\documentclass[a4paper,aps,pra,showpacs,superscriptaddress,twocolumn]{revtex4-1}
\usepackage[utf8]{inputenc}
\usepackage[T1]{fontenc}

\usepackage{hyperref}
\usepackage{color}
\usepackage{amsfonts,dsfont,mathrsfs,amsmath,amssymb,bm,bbm}
\usepackage[english]{babel}
\usepackage[babel]{microtype}  
\usepackage{graphicx}
\usepackage{subfigure}
\usepackage{braket}

\newcommand{\myref}[1]{Equation \eqref{#1}}
\newcommand{\mycite}[1]{ref. \cite{#1}}
\newcommand{\mycites}[1]{refs. \cite{#1}}
\newcommand{\myfig}[1]{Figure \ref{#1}}
\newcommand{\bgfg}{\begin{figure}}
\newcommand{\edfg}{\end{figure}}

\newcommand{\F}{\mathbb{F}_{2}}
\newcommand{\gf}[1]{\mathbb{F}_{#1}}
\newcommand{\gfx}[1]{\mathbb{F}_{#1}[x]}

\begin{document}

\title{Quantum-accelerated algorithms for generating random primitive polynomials over finite fields}

\author{Shan Huang}
\affiliation{National Laboratory of Solid State Microstructures and School of Physics, Collaborative Innovation Center of Advanced Microstructures, Nanjing University, Nanjing 210093, China}
\affiliation{Institute for Brain Sciences and Kuang Yaming Honors School, Nanjing University, Nanjing 210023,
China}

\author{Hua-Lei Yin}
\email{hlyin@nju.edu.cn}
\affiliation{National Laboratory of Solid State Microstructures and School of Physics, Collaborative Innovation Center of Advanced Microstructures, Nanjing University, Nanjing 210093, China}

\author{Zeng-Bing Chen}
\email{zbchen@nju.edu.cn}
\affiliation{National Laboratory of Solid State Microstructures and School of Physics, Collaborative Innovation Center of Advanced Microstructures, Nanjing University, Nanjing 210093, China}

\author{Shengjun Wu}
\email{sjwu@nju.edu.cn}
\affiliation{National Laboratory of Solid State Microstructures and School of Physics, Collaborative Innovation Center of Advanced Microstructures, Nanjing University, Nanjing 210093, China}
\affiliation{Institute for Brain Sciences and Kuang Yaming Honors School, Nanjing University, Nanjing 210023,
China}

\begin{abstract}
Primitive polynomials over finite fields are crucial resources with broad applications across various domains in computer science, including classical pseudo-random number generation, coding theory, and post-quantum cryptography. Nevertheless, the pursuit of an efficient classical algorithm for generating random primitive polynomials over finite fields remains an ongoing challenge.  In this work, we show how this problem can be solved efficiently with the help of quantum computers.  Moreover, designs of the specific quantum circuits to implement them are also presented. Our research paves the way for the rapid and real-time generation of random primitive polynomials in diverse quantum communication and computation applications.
\end{abstract}

\maketitle

\section{Introduction}

The concept of quantum computation dates back to Feynman, who pointed out in 1982 that a suitable quantum machine should be able to outperform its classical counterparts in simulating quantum systems \cite{Feynman94,Feynman95}. Although it is still an ongoing research to explore the power of quantum computers, quantum algorithms are known to be with unparalleled superiority over the classical ones in several computational tasks \cite{Deutsch1992,Grover97,Shor94}, among which Shor's algorithms for integer factorization and discrete logarithm problem achieve an exponential acceleration over all existing classical ones \cite{Shor94}. Shor's algorithms thus make those currently widely used public key cryptosystems based on the difficulty of factoring large integers or discrete logarithm problems, e.g., RSA and ECC,  completely insecure under quantum attack. This has raised increasing attention among researchers in developing post-quantum cryptography (PQC) \cite{bernstein2017post}, namely,  cryptosystems that are resistant against the attack of quantum computers. In the upcoming era of quantum computing, PQC is essential for the security of numerous common applications in our daily lives, such as online communication, implantable or wearable devices \cite{sarker2022,mozaffari2015,mozaffari2011,Sarker2022efficient} and the software industry\cite{awan2022}.

Irreducible polynomials over finite fields, particularly primitive ones, play an essential role in PQC. For example, they are frequently used to construct error-correcting codes \cite{kasami69,goppa1970,macwilliams1977,roth06,lidl1994introduction,cohen1993primitive,cardell2013}, which constitute the foundation of many code-based public key cryptosystems \cite{mceliece1978,niederreiter1986,overbeck09,kuznetsov17}, including code-based hash functions \cite{grossman04,applebaum2017low,brakerski2019worst,yu2019collision}, code-based zero-knowledge proof protocols \cite{aguilar11,cayrel2010} and code-based identification and signature schemes \cite{kuznetsov19,gorbenko16,stern1993,veron1997,song2020}.  Code-based cryptography is recently recognized as one of the four most promising avenues to PQC (see the review \cite{bernstein2017post} and references therein for details).

Irreducible polynomials are also the characteristic polynomials of linear feedback shift registers (LFSR) for generating pseudo-random sequences \cite{golomb81}, and those sequences generated by primitive polynomials are proven to be with the best randomness properties \cite{golomb81,blackburn1996,macwilliams1976,fredricsson1975,mitra08}.
In Yuen's protocol (Y00) of keyed communication in quantum noise \cite{yuen03,yuen04,Barbosa03,Yuen09,chuang05}, LFSR-based pseudo-random numbers are generated from the shared secret keys between two communication parties, which are subsequently utilized to encode data bit information into proper quantum states or to select the optimal measurement for decoding the data bit information from quantum states. Y00-type protocols are famous for highly efficient in communication tasks while keeping computational security under quantum attack.

Despite the broad applications of primitive polynomials in computer science and PQC,  existing classical algorithms for generating random degree $n$ primitive polynomials over a finite field containing $q$ elements, denoted $\gf{q}$, rely on the prime factorization of  $q^n-1$ to be efficiently implemented \cite{shoup90,rieke98,hansen92,vijayarangan05,shparlinski96}, which greatly narrows their practical applications. Fortunately, integer factorization is efficiently solvable on quantum computers. Drawing inspiration from Simon's algorithm \cite{simon1997}, Shor proposed a quantum algorithm \cite{Shor94} for factoring large integers, say $N$, in time O($(\log N)^3$).

However, we emphasize that there exists more straightforward and simpler algorithms for directly testing primitivity, instead of utilizing Shor's algorithm to factorize $q^n-1$ and subsequently generating random primitive polynomials over $\gf{q}$ on a classical computer.  Shor's approach is to reduce the problem of factoring to the order-finding of a random integer $1<y<N$ that is prime to $N$, i.e., finding the smallest positive integer $r$ satisfying $y^r=1\mod N$. As we will see in the following section, the problem of primitivity test of a degree $n$ polynomial over $\gf{q}$ can be reduced to the order-finding of an element of a cyclic group with $q^n-1$ elements.  Leveraging the specific properties of a cyclic group and the fact that, unlike factoring, the primitivity test is essentially a binary decision problem, considerable simplification can be achieved.
Moreover, it is also of great practical significance to explore novel algorithms that, compared to order-finding, exhibit higher noise robustness while maintaining relatively low overall complexity.

This article is structured as follows.  In Sec. \ref{pre}, we provide background knowledge about primitive polynomials over finite fields and the criterion for testing primitivity.  In Sec. \ref{orderfinding}, we introduce a simplified order-finding algorithm to test the primitivity of random polynomials over finite fields. Further, we propose in Sec. \ref{generate} another algorithm for primitivity test. It is probabilistic but more robust against noise. In Sec. \ref{complexity}, we analyze the complexity of our algorithms. In Sec. \ref{discussion},  we make some further discussions about our algorithms. Finally, we draw a brief conclusion in Sec. \ref{conclusion}.

\section{\label{pre}Preliminary}

The finite field $\gf{q}$ consists of $q$ elements associated with two binary operations: addition and multiplication.  $\gf{q}$ is an additive group under addition, and the $q-1$ nonzero elements of  it  form the cyclic group $F_q^\times$ under multiplication. An example of finite field is the congruence classes of integers modulo some prime integer $p$. The requirement that $p$ being prime is necessary for the nonzero congruence classes to be multiplicative invertible and thus form a cyclic group.  Another example is, with $\gfx{q}$ denoting the set of all polynomials with coefficients lying in $\gf{q}$, the congruence classes of polynomials in $\gfx{q}$ modulo some degree $n$ polynomial $p(x)$ that cannot be factored into polynomials of smaller degrees dividing $n$. In this case, $p(x)$ is said to be irreducible and the corresponding field, denoted $\gf{q^n}$, is called a degree $n$ extension of $\gf{q}$. $\gf{q^n}$ contains $q^n$ elements, each corresponding to a unique congruence class represented by a polynomial over $\gf{q}$ of degree smaller than $n$. Degree $n$ primitive polynomials over $\gf{q}$ are irreducible polynomials over $\gf{q}$ that have $n$ distinctive generators (primitive elements) of the cyclic group $\gf{q^n}^\times$ as their roots.

\subsection{Criterion for primitivity}

Observe that for any degree $n$ irreducible polynomial $p(x)$ over $\gf{q}$, the congruence classes $x^{q^k}\mod p(x)$ $(k=0,\cdots,n-1)$ are solutions to $p(x)=0$ since $0\mod p(x)=[p(x)]^{q^k}\mod p(x)=p(x^{q^k}\mod p(x))$. Therefore, $p(x)$ is primitive over $\gf{q}$ if and only if $x\mod p(x)$ is a generator of $\gf{q^n}^\times$, or in other words, the cyclic group $G_x=\{x\mod p(x), x^2\mod p(x),\cdots,x^{r-1}\mod p(x)\}$ generated by $x\mod p(x)$ is exactly the multiplicative group $\gf{q^n}^\times$. In this case, obviously it holds that  $x^{q^n-1}=1\mod p(x)$. But the opposite is not necessarily true. From $x^{q^n-1}=1\mod p(x)$ we can only conclude that powers of $x$ are not divisible by $p(x)$ and the order of $x\mod p(x)$---the minimum positive integer $r$ satisfying $[x\mod p(x)]^r=1\mod p(x)$---must be a divisor of $q^n-1$, which means $p(x)$ is irreducible.  For example, $x^4+x^3+x^2+x+1$ is irreducible but not primitive over $\F$, as it can be easily checked that the order of $x$ modulo it is 5, a divisor of $2^4-1$. Thus, given a degree $n$ irreducible polynomial $p(x)$ over $\gf{q}$, a further check on whether the order of $x\mod p(x)$ is $q^n-1$ or one of its nontrivial factors instead is necessary for testing the primitivity of $p(x)$.

The conditions for a degree $n$ polynomial $p(x)$ to be primitive over $\gf{q}$ can be concluded as follows:
\begin{equation}\left\{\begin{aligned}
	&(a).\ \ x^{q^n-1}=1\mod p(x).\\
	&(b).\ \ x^{q^t-1}\neq1\mod p(x),\\
	&\text{for any prime factor $t$ of $n$}.\\
	&(c).\ \ x^\frac{q^n-1}{d}\neq 1\mod p(x),\\
	&\text{for any prime factor $d$ of $q^n-1$}.
\end{aligned}\right.\label{testcond}\end{equation}
 It is worth mentioning that in \myref{testcond}, $(a)$ combined with $(c)$ is sufficient for $p(x)$ to be primitive, while $(a)$ combined with $(b)$ is the criterion for irreducibility \cite{rabin1980} and can be tested efficiently on a classical computer \cite{brent1978fmc, huang1998fmc,kedlaya2008fmc}.
 To reduce quantum resources, we can check $(a)$ and $(b)$ on a classical computer before moving on to $(c)$, since only approximately $1/n$ of degree $n$ polynomials over $\gf{q}$ are irreducible.

We may also generate degree $n$  primitive polynomials over $\gf{q}$ from primitive elements of the extension field $\gf{q^n}$ (i.e., generators of $\gf{q^n}^\times$) based on the following criterion:
\begin{equation}\left\{\begin{aligned}
	&(a).\ \ \alpha^{q^i}\neq\alpha^{q^j},\ (\forall\ 0\leq i\neq j\leq n-1)\\
	&(b).\ \ \alpha^\frac{q^n-1}{d}\neq 1,\ \alpha\neq0, \\
	&\text{for any prime factor $d$ of $q^n-1$}.
\end{aligned}\right.\label{testcond1}\end{equation}
The second condition in \myref{testcond1} is necessary and sufficient for $\alpha\in\gf{q^n}$ to be a primitive element, while the first one promises $\alpha$ to be a root of some degree $n$ irreducible polynomial over $\gf{q}$ and is efficient testable on a classical computer.  To reduce quantum cost, again, we test $(a)$ of \myref{testcond1} on a classical computer before moving on to $(b)$.

\subsection{\label{representation}Representation of finite field and modular multiplication}

 Consider now a degree $n$ irreducible polynomial $p(x)$ over finite field $\gf{q}$ and the associated extension field $\gf{q^n}$ which consists of the $q^n$ congruence classes of polynomials over $\gf{q}$ modulo $p(x)$.  For each element $s(x)\mod p(x)$ of $\gf{q^n}$, where $s(x)=s_{n-1}x^{n-1}+\cdots+s_1x+s_0$ can be any polynomial over $\gf{q}$ of degree smaller than $n$, we utilize the quantum state  $\ket{s(x)\mod p(x)}=\ket{s_{n-1}}\otimes\cdots \ket{s_1}\otimes \ket{s_0}=\ket{s_{n-1}s_{n-2}\cdots s_0}$ to represent it on a quantum computer. For example, consider the case $q=2$ and $p(x)=x^3+x+1$, the three-qubit state $\ket{011}$ represents $x+1\mod x^3+x+1$, which is an element of $\gf{2^3}$. Under this notation, we can represent the $q^n$ elements of $\gf{q^n}$ with exactly $n\log q$ qubits when $q$ is a power of 2.

Let $G_x=\{1\mod p(x),\ x\mod p(x),\cdots,$ $ x^{r-1}\mod p(x)\}$ be the order $r$ ($r$ divides $q^n-1$) cyclic group generated by $x\mod p(x)$ and $U_x$ an unitary representation of  $x\mod p(x)\in G_x$, then we have
\begin{equation}
U_{x}\ket{s(x)\mod p(x)}=\ket{xs(x)\mod p(x)}\label{unitary}.
\end{equation}

Here a more intuitive way of understanding $U_x$  may be, instead of unitary representation of  $x\mod p(x)$,  a quantum multiplier over $\gf{q^n}$ that multiplies $s(x)\mod p(x)$ by $x\mod p(x)$ for any $s(x)\mod p(x)\in \gf{q^n}$. To build the quantum circuit for implementing $U_x$, without loss of generality, let us take $q=2$ as an example. Suppose now $p(x)=x^n+a_{n-1}x^{n-1}+\cdots+a_2x^2+a_1x+1$, observe that $x^n\mod p(x) =a_{n-1}x^{n-1}+\cdots a_1x+1\mod p(x)$, hence $U_x$ is essentially the following transformation
\bgfg[b]
	\centering
	\includegraphics[width=8.6cm]{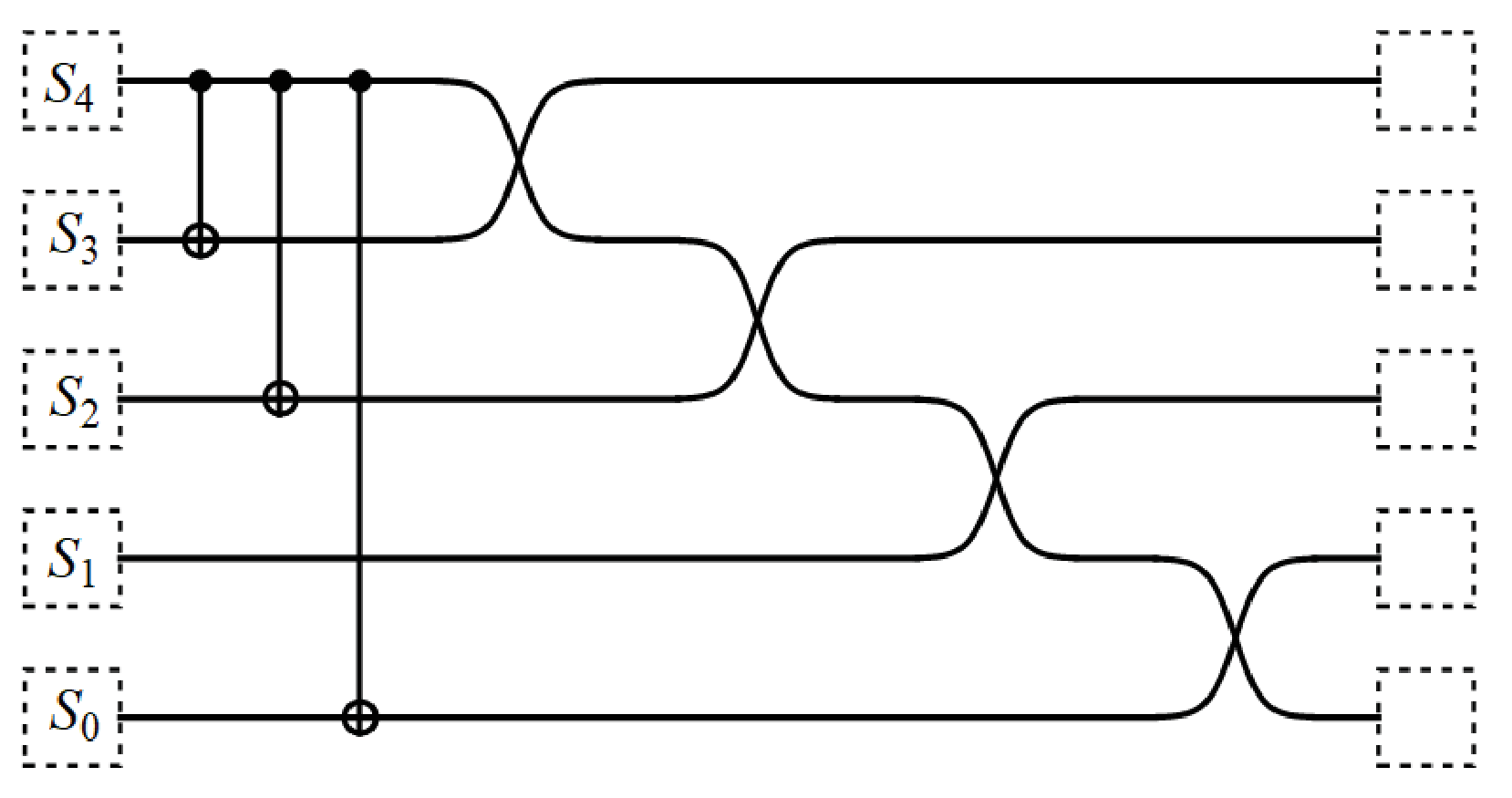}\\
	\caption{The quantum circuit for implementing $U_{x}$ when $p(x)=x^5+x^4+x^3+x+1$ is a polynomial over $\gf{2}$.}\label{ucircuit}
\edfg
\begin{align}
	&s(x)\mod p(x)\longrightarrow xs(x)\mod p(x)=\nonumber\\
	&\big[(s_{n-2}\oplus a_{n-1}s_{n-1})x^{n-1}+\cdots+(s_0\oplus a_1s_{n-1})x\nonumber\\
	&+s_{n-1}\big]\mod p(x)\label{xsx}
\end{align}

This transformation can be factored into two parts. First, for those $0\leq k\leq n-2$, flip the $k$\text{-}th qubit if  $a_{k+1}s_{n-1}$=1, this can be done by $|p(x)|-2$ controlled-NOT (CNOT) gates, where $|p(x)|$ is the number of nonzero coefficients of $p(x)$. Second, shift the state of the $n$ qubits cyclically as follows
$$\ket{s_{n-1}s_{n-2}\cdots s_1s_0}\longrightarrow\ket{s_{n-2}\cdots s_1s_0s_{n-1}}.$$
This shift can be realized via $n-1$ consecutive SWAP operations on pairs of neighboring qubits. Thus the unitary $U_{x}$ described in \myref{unitary} can be factored into a series of 2-qubit unitaries, including no more than $n-1$ CNOT gates and followed by $n-1$ SWAP gates. As an example, consider the case $n=5$ and $p(x)=x^5+x^4+x^3+x+1$,  a quantum circuit for the implementation of $U_{x}$ is illustrated in \myfig{ucircuit}.

Our algorithms require also the multipliers $\{U_x^{2^1},\ U_x^{2^2}$, $\cdots,\ U_x^{2^{n-1}}\}$, a combination of which would enable us to perform the $j$th power transformation $U_x^j$ for any $1\leq j\leq 2^n-1$ (note that according to $(a)$ of Equation \eqref{testcond}, $U_x^{2^n}=U_x$ holds whenever $p(x)$ is reducible). To construct these unitaries we can firstly compute $x^{2^k}\mod p(x)$ for $k=1,2,\cdots,n$ on a classical computer, based on which $U_{x}^{2^k}=U_{x^{2^k}}$ (or $U^{2^k}$ for simplicity) can be constructed from $x^{2^k}\mod p(x)$ quite in the way we construct $U_x$ from $x\mod p(x)$.

\section{\label{algorithm}Algorithms}

In this section, we will delve into two distinct approaches for generating primitive polynomials over finite fields. The first method involves testing the primitivity of randomly generated polynomials using Equation \eqref{testcond}. On the other hand, the second method utilizes Equation \eqref{testcond1} to randomly generate a primitive element (a generator of $\gf{q^n}^\times$), denoted as $\alpha\in\gf{q^n}$, on a quantum computer. Subsequently, classical algorithms for computing the minimal polynomial \cite{massey69, shoup99} can be applied to obtain a degree $n$ primitive polynomial such that $\alpha$ is a root of it.

\subsection{\label{orderfinding}Primitivity test by order-finding}

Shor's order-finding algorithm  \cite{Shor94} can be harnessed (see also Simon's algorithm \cite{simon1997} and the lecture \cite{wright}) to solve our problem by finding the order $r$ of $x\mod p(x)$ directly, whereas it can be considerably simplified considering that, if $p(x)$ is an irreducible polynomial over $\gf{q}$, $x\mod p(x)$ must generate a subgroup of $\gf{q^n}^\times$ with an order dividing $q^n-1$. Next, without loss of generality, we will take the random generation of primitive polynomials over $\F$ as an example.

Our algorithm for testing the primitivity of a degree $n$ polynomial $p(x)$ over $\F$ goes as follows (see \myfig{orderfind} for the quantum circuit).

\emph{1.} Initialize a register with $n$ qubits into the state $\ket{\mathbf{0}}=\ket{0}^{\otimes n}$ and prepare in another register the $n$-qubit state $\ket{\mathbf{1}}=\ket{0}^{\otimes n-1}\otimes\ket{1}$ to represent $1\mod p(x)$.

\emph{ 2.} Apply the Hardmard transform to each of the $n$ qubits in the first register to obtain the state
$$H^{\otimes n}\ket{\mathbf{0}}=\left(\frac{\ket{0}+\ket{1}}{\sqrt{2}}\right)^{\otimes n}=\frac{1}{\sqrt{N+1}}\sum_{j=0}^{N}\ket{j},$$
where  $N=2^n-1$ and $\ket{j}$ is the $n$-qubit binary representation of the integer $j$.

\emph{3.} Prepare an ancillary qubit in the state $\ket{0}_{\rm anc}$ and, controlled by the qubits in register 1, make a multiple control NOT operation  (see Appendix. \ref{Togate}) on it. Then the remaining state would be 
$$\frac{1}{\sqrt{N+1}}\big(\sum_{j=0}^{N-1}\ket{j}\ket{0}_{\rm anc}+\ket{N}\ket{1}_{\rm anc}\big).$$

By measuring the ancillary qubit we'll succeed with probability $N/(N+1)$ in preparing  in register 1 a uniform superposition $\frac{1}{\sqrt{N}}\sum_{j=0}^{N-1}\ket{j}$.

\emph{4.} Construct the unitaries $\{U^{2^k}\}$ $(1\leq k\leq n-1)$ in a way as discussed in Sec. \ref{representation} and apply $U^{2^k}$ to the qubits in register 2 if the $k$-th qubit in register 1 is $\ket{1}$. So the transformation of the whole system is

$\begin{aligned}
&\frac{1}{\sqrt{N}}\sum_{j=0}^{N-1}\ket{j}\ket{\mathbf{1}}\longrightarrow\frac{1}{\sqrt{N}}\sum_{j=0}^{N-1}\ket{j}(U^j\ket{\mathbf{1}})\\
=&\frac{1}{\sqrt{N}}\sum_{j=0}^{N-1}\ket{j}\ket{x^j \mod p(x)}.
\end{aligned}$\\

\emph{5.} Perform an approximated inverse Fourier transform over $\mathbb{Z}_{2^n-1}$ (see Appendix. \ref{QFT}) on the first register and then measure the qubits in it
\begin{align}
	&\frac{1}{\sqrt{N}}\sum_{j=0}^{N-1}\ket{j}\ket{x^j}=\frac{1}{\sqrt{N}}\sum_{a=0}^{\frac{N}{r}-1}\sum_{b=0}^{r-1}\ket{ar+b}\ket{x^b}\nonumber\\
	&\stackrel{FT^+}{\longrightarrow}
	\frac{1}{N}\sum_{b=0}^{r-1}\sum_{l=0}^{N-1}\left(\sum_{a=0}^{\frac{N}{r}-1}e^\frac{i2\pi arl}{N}\right)e^\frac{i2\pi bl}{N} \ket{l}\ket{x^b}.\label{fourier}
\end{align}

\bgfg[t]
\centering
\includegraphics[width=8.6cm]{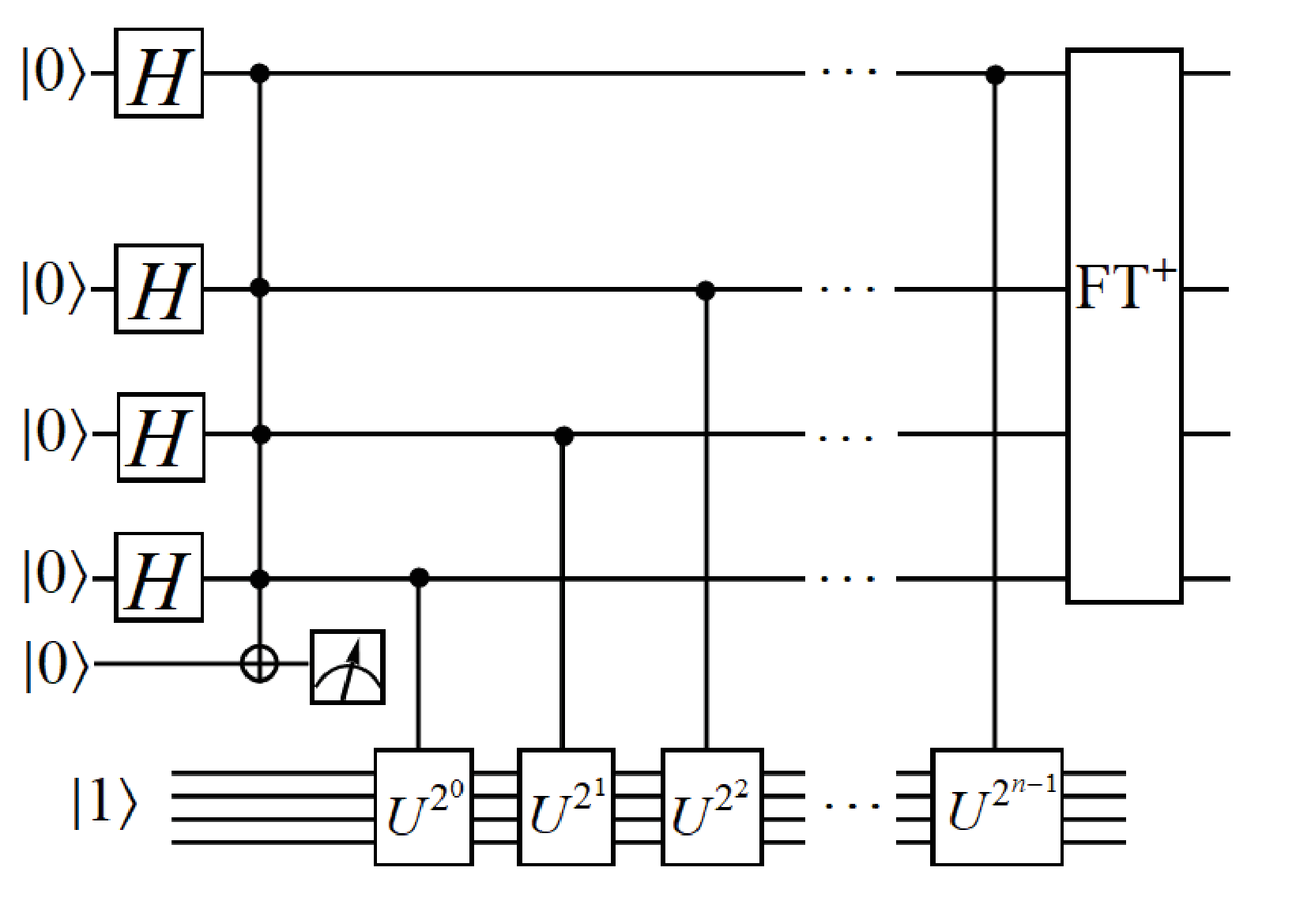}
\caption{The quantum circuit for testing the primitivity of polynomials over $\F$ by subgroup order-finding.}\label{orderfind}
\edfg

As $r$ must be a divisor of $N=2^n-1$,  according to \myref{fourier} the probability of obtaining the integer $l$ is
\begin{align}
	p(l)=&\frac{r}{N^2}\Big|\sum_{a=0}^{N/r-1}e^\frac{i2\pi arl}{N}\Big|^2\nonumber\\
	=&\left\{\begin{aligned}&\frac{1}{r},\  \ l=0,\frac{N}{r},\frac{2N}{r}\cdots,\frac{(r-1)N}{r};\\
		&0,\  \text{$l$ is not a multiple of $\frac{N}{r}$}.
	\end{aligned}\right.
\end{align}
Hence there are $r$ possible results with equal probability $1/r$,  each being a multiple of $\frac{N}{r}$.

Suppose we have repeated this algorithm  $L$ $(L\geq1)$ times and the measurement results are respectively $l_1,l_2,\cdots,l_L$.  We replace those measurement results being 0 by $N$ and then run the Euclidean algorithm for computing the great common divisor (GCD) on a classical computer. If $GCD(N,l_1,l_2,\cdots,l_L)=1$, we immediately know that $\frac{N}{r}=1$ and, consequently, $p(x)$ is primitive. While if $GCD(N,l_1,l_2,\cdots,l_L)=g>1$ is a divisor of $N$, we go on to check on a classical computer whether $r=N/g$. We declare that $p(x)$ is not primitive if the answer is yes, or we need to repeat the algorithm to obtain more integers $l_{L+1}, l_{L+2}, \cdots$ and update the value of $g$ to $GCD(g,\ l_{L+1})\cdots$. This process proceeds until we get $g=1$ ($r=N$; primitive) or $g>1$ and $r=N/g$ ($r<N$; not primitive).

\bgfg[t]
\centering
\includegraphics[width=8.6cm]{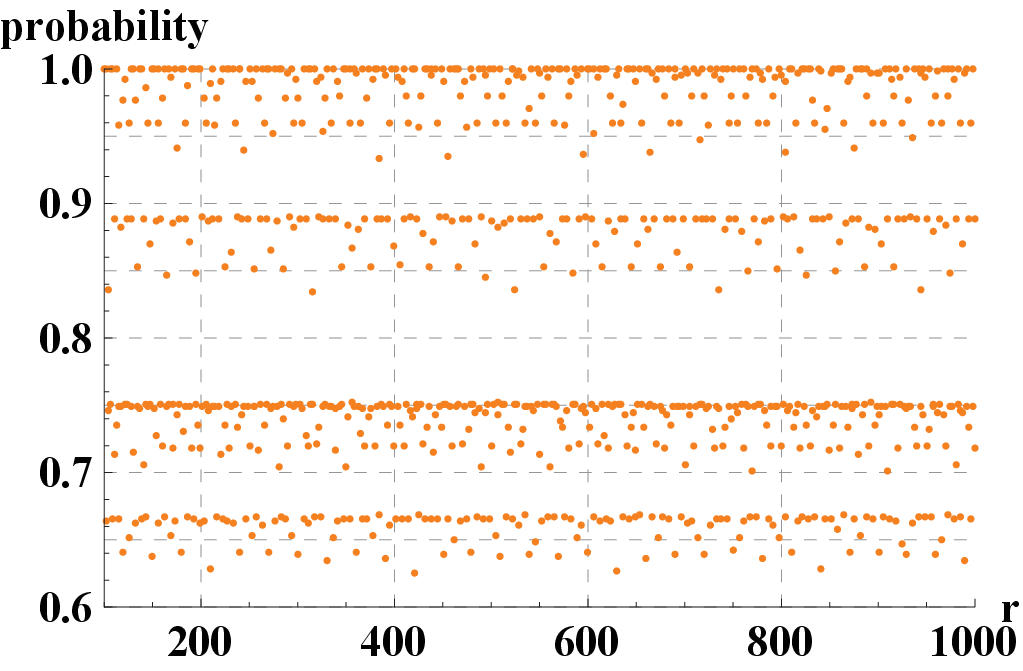}
\caption{The probability that  $r$ $(100\leq r\leq 1000)$ is prime to the GCD of two random integers in $[1, r]$ .}\label{gcd}
\edfg

Now the main question to determine the probability that we successfully obtain $\frac{N}{r}$.  Numerical results indicate that this probability is sensitive to $r$ and the case $L=2$ is depicted in \myfig{gcd}. As shown, the success probability is generally much larger than $0.6$. For more general results, it's complicated to draw useful conclusions from $GCD(N,l_1,l_2,\ \cdots,l_L)$ directly. To simplify our analysis,  let $\mathscr{P}_r(L)$ denote the probability that $GCD(l_1,l_2,$ $\cdots,l_L)=\frac{2^n-1}{r}$ for any $L\geq 2$. This probability is obviously equal to that of $L$ random integers from $\{1,2,\cdots,r\}$ being co-prime and it won't be larger than the success probability of obtaining $\frac{N}{r}$ by repeating our algorithm $L$ times.

\begin{figure}[t]
	\centering
	\includegraphics[width=8.6cm]{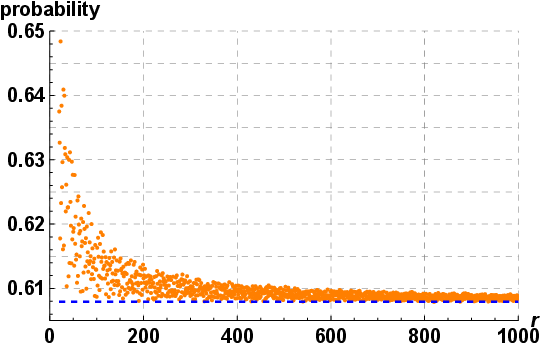}
	\caption{The probability that two random positive integers in $[1, r]$ $(20\leq r\leq 1000)$ are co-prime.}\label{proba}
\end{figure}

In the case $L=2$ there is \cite{Mertens1874}
\begin{equation}
\mathscr{P}_r(2)=\frac{1}{r^2}\Big(2\sum_{i=0}^r\phi(r)-1\Big)= \frac{6}{\pi^2}+O\Big(\frac{\log r}{r}\Big),\label{probre2}
\end{equation}
where $\phi$ is the Euler's totient function, referring to the number of integers in the range $[1,r]$ that are prime to $r$. Roughly speaking, as $r$ increases, $\mathscr{P}_r(2)$ decreases and eventually approaches the asymptotic limit of $\frac{6}{\pi^2}\approx0.608$ for infinitely large $r$. Some numerical results about $\mathscr{P}_r(2)$  for $100\leq r\leq 1000$ are presented in \myfig{proba},  where we can see $\mathscr{P}_r(2)$ is quite close to $\frac{6}{\pi^2}$ for $r\approx 1000$.  Consequently, for any degree $n>10$ ( $2^n-1>1000$) non-primitive polynomial $p(x)$ the probability that we are able to assert $p(x)$ is not primitive with certainty by repeating the algorithm only two times is approximately equal to or larger than $\frac{6}{\pi^2}$. Correspondingly, the probability that we fail to assert the primitivity of a primitive polynomial is approximately equal to or smaller than $1-\frac{6}{\pi^2}$.

For $L>2$, note that $L$ integers are co-prime whenever two of them are co-prime, we have $\mathscr{P}_r(L)\approx\mathscr{P}_{1000}(L)\geq1-(1-\frac{6}{\pi^2})^{L/2}$. Interestingly,  our numerical results imply that $\mathscr{P}_r(L)$ is almost independent of $r$ when $r$ is sufficiently large $(r\geq 1000)$, and approaches 1 as $L$ tends to infinity. Table \ref{table} presents the values of $\mathscr{P}{1023}(L)$ for $2\leq L\leq 6$.

\begin{table}[htbp]
\centering
\setlength\tabcolsep{2.2pt}
\begin{tabular}{|c|c|c|c|c|c|}
	\hline
	L&2&3&4&5&6\\
	\hline
	probability&0.608&0.832&0.924&0.9644&0.983\\
	\hline
\end{tabular}
\caption{The probability that L $(2\leq L\leq6)$ random integers in  $[1, 2^{10}-1]$ are co-prime.}\label{table}
\end{table}

\subsection{\label{generate}Random search for primitive element}

The main idea of this algorithm is to prepare an approximated uniform superposition state of  all group elements of  the cyclic group $G_\alpha=\{1,\alpha,\cdots,\alpha^{r-1}\}$ generated by a random element $\alpha\in\gf{q^n}$, then check if it is invariant under proper permutations on $\gf{q^n}$.

Suppose $p(x)$ is a degree $n$ irreducible polynomial over $\gf{q}$ and $\gf{q^n}$ is an extension field defined by $p(x)$. We represent $\gf{q}$ and $\gf{q^n}$ ($q=p^m$, $p$ is prime) with $m\lceil\log p\rceil$ and $nm\lceil\log p\rceil$ qubits respectively. Let $A:\ \gf{q^n}\rightarrow\gf{q^n}$ be the Fibonacci automorphism which maps $\alpha$ to $\alpha^q$. For example, consider $s(x)=s_{n-1}x^{n-1}+\cdots+s_1x+s_0$ $( s_i\in\gf{q})$, then
\begin{align}
&A:\ s(x)\mod p(x)\rightarrow\left[s(x)\mod p(x)\right]^q\nonumber\\
=& s_{n-1}x^{q(n-1)}+\cdots+s_1x^q+s_0\mod p(x)\label{QFA}
\end{align}
With $U_A$ being the unitary that transforms the quantum state  $\ket{\alpha}$ to $\ket{\alpha^q}$ for any $\alpha\in\gf{q^n}$,  we random search for primitive elements of $\gf{q^n}$ as follows.

\emph{1}. Random select an element $\alpha\in\gf{q^n}$ satisfying  $(a)$ and $(b)$ of  \myref{testcond1} for any $d\in[2,\ D]$,  where $D\geq2$ is an integer. $D$ should be chosen sufficiently large to ensure our average success probability in each round of this algorithm, but its value is intimately related to the amount of quantum computation involved in each round.

\emph{2}.  Let $\ket{\psi_k(\alpha)}=\ket{1}+\ket{\alpha}+\cdots+\ket{\alpha^{q^{k+1}-1}}$ and  $\ket{\gamma_k(\alpha)}=U_A^k\ket{\psi_0(\alpha)}=\ket{1}+\ket{\alpha^{q^k}}+\ket{\alpha^{2q^k}}\cdots+\ket{\alpha^{(q-1)q^k}}\ (0\leq k\leq n-1)$ be two $n\lceil \log q\rceil$-qubit state. Prepare the state $\ket{\gamma_0}\otimes\ket{\psi_0}\otimes\ket{\mathbf{0}}$, where  $\ket{\mathbf{0}}=\ket{0}^{\otimes n\lceil\log q\rceil}$, $\ket{\psi_0}=\ket{\gamma_0}=\ket{1}+\ket{\alpha}+\cdots+\ket{\alpha^{q-1}}$ and the normalization factor is ignored.

\emph{3}. Let $Mul:\ \ket{\alpha}\otimes\ket{\beta}\otimes\ket{\mathbf{0}}\rightarrow\ket{\alpha}\otimes\ket{\beta}\otimes\ket{\alpha\beta}$ be a reversible quantum general  multiplier over $\gf{q^n}$. It can be easily checked that
$Mul\ (\ \ket{\gamma_{k+1}}\otimes\ket{\psi_k}\otimes\ket{\mathbf{0}})=\ket{\gamma_{k+1}}\otimes\ket{\psi_k}\otimes\ket{\psi_{k+1}}$. Apply $U_A$ to $\ket{\gamma_0}$ and compute $\ket{\psi_1}$ from $\ket{\gamma_1}\otimes\ket{\psi_0}\otimes\ket{\mathbf{0}}$, repeat this procedure to obtain $\ket{\gamma_2}\otimes\ket{\psi_2},\cdots, \ket{\gamma_{n-1}}\otimes\ket{\psi_{n-1}}$ consecutively.
\begin{align}\label{psi}
&\ket{\gamma_1}\otimes\ket{\psi_0}\otimes\ket{\mathbf{0}}\longrightarrow \ket{\gamma_1}\otimes\ket{\psi_0}\otimes\ket{\psi_1},\nonumber\\
&\ket{\psi_1}=\ket{1}+\ket{\alpha}+\cdots+\ket{\alpha^{q^2-1}}\nonumber\\
&\ket{\gamma_2}\otimes\ket{\psi_1}\otimes\ket{\mathbf{0}}\longrightarrow \ket{\gamma_2}\otimes\ket{\psi_1}\otimes\ket{\psi_2},\nonumber\\
&\ket{\psi_2}=\ket{1}+\ket{\alpha}+\cdots+\ket{\alpha^{q^3-1}}\nonumber\\
&\cdots\cdots\nonumber\\
&\ket{\psi_{n-1}(\alpha)}=\ket{1}+\ket{\alpha}+\cdots+\ket{\alpha^{q^n-1}}
\end{align}

If $\alpha$ is a primitive element, then $\ket{\psi_{n-1}(\alpha)}$ is nearly a uniform superposition of all nonzero elements in $\gf{q^n}$, effectively acting as an invariant state with respect to any permutation on $\gf{q^n}$. The permutations we consider here are the additive shifts $P_w: \ket{\alpha^j}\rightarrow\ket{\alpha^j+w} \ (\alpha^j\in G _\alpha,\ w\in\gf{q^n}^\times)$. In polynomial representation, an additive shift $P_w$ on $\gf{q^n}$ is nothing but additions between $w$ and the congruence classes of polynomials over $\gf{q}$ modulo $f(x)$, which consists of $n$ independent additions on $\gf{q}$, each with gate complexity $O(\log q)$ corresponding to monomials of different degrees.

\emph{4}.  Initialize a qubit in the state $(\ket{0}_c+\ket{1}_c)/\sqrt{2}$, controlled by which  perform the $P_1$ transform on  $\ket{\psi_{n-1}}$ to obtain
$\left(\ket{0}_c\otimes\ket{\psi_{n-1}}+\ket{1}_c\otimes P_1\ket{\psi_{n-1}}\right)/\sqrt{2}$,  then apply a Hadamard transform to the control qubit and measure it.
\begin{align}
\ket{\psi}=&\frac{1}{2}\ket{0}_c\otimes\left(\ket{\psi_{n-1}}+P_1\ket{\psi_{n-1}}\right)\nonumber\\
+&\frac{1}{2}\ket{1}_c\otimes \left(\ket{\psi_{n-1}}-P_1\ket{\psi_{n-1}}\right)\label{classify}
\end{align}
For primitive $\alpha$, the measurement result would be 0 with probability $1-2/(q^n+2)$, or we will obtain 1 with probability about 1/2.
 \begin{equation}
\alpha^a+\alpha^b=w\ (0\leq a,b\leq r-1;  w\in\gf{q^n}^\times)\label{cyclotomy}
\end{equation}
To see why is this, suppose now $w=1$, 
then the number of solutions $K$ to \myref{cyclotomy} in an order $r$ subgroup  of $\gf{q^n}^\times$ generated by $\alpha$ is bounded by the following \cite{WILSON1972,betsumiya13},
\begin{align*}
\Big |K-\frac{r^2(q^n-2)}{q^{2n}}\Big|<\sqrt{q^n-1}.
\end{align*}
When $\sqrt{q^n}\ll r\ll q^n$ we immediately have $K/r\ll1$, more over, for a random element $\alpha\in\gf{q^n}^\times$ of order $r$ the expected value of $K$ is about $r^2/q^{n}$, leading to $\bra{\psi_{n-1}}P_1\ket{\psi_{n-1}}=K/r\approx r/q^{n}<1/D$.  Hence $G_\alpha$ is generally far from closed under $P_1$ transform, which is the foundation of our technique in stage 4.
However, when $r+1$ is a divisor of $q^n$ there are $r-1$ solutions to  \myref{cyclotomy} with $w=1$. Although this special case can be excluded easily via classical computers, we don't know if there are more similar situations. The following stage aims to exclude the above ambiguity.

 \emph{5}. We progress on with the state, denoted by $\ket{\psi^0}$, obtained by measuring the control qubit in $\ket{0}_c$. Initial another qubit in $(\ket{0}_c+\ket{1}_c)/\sqrt{2}$ to control a random additive shifts $P_w$ on $\ket{\psi^0}$, then perform a Hadamard transform on the control qubit and measure it.

In the case $q$ is odd, for any odd $r$ there is $-1\notin G_\alpha$, consequently, the addition between arbitrary two elements of $G_\alpha$ cannot be 0.  The average number of solutions to \myref{cyclotomy}  is $r^2/(q^n-1)$ when $w$ is random. For even $r$,  $-1\in G_\alpha$ and for each element $\alpha^a$ in $G_\alpha$ there exists one and only one element $\alpha^b\in G_\alpha$ such that $\alpha^a+\alpha^b=0$. Now the average number of solutions to \myref{cyclotomy} becomes $r(r-1)/(q^n-1)$ when $w$ is random. So approximately $\frac{r}{q^n-1}$  elements of $G_\alpha$ stay in $G_\alpha$ after a random additive shift $P_w$, and we have $\bra{\psi^0}P_w\ket{\psi^0}\approx r/(q^{n}-1)<\frac{1}{D}$.  Meanwhile, we can repeat the stage 5 more times $(<\log D)$, during which just one result 1 would imply non-primitivity. If all the measurement results are 0 we may repeat the whole algorithm to increase our confidence level, therefore by repeating this algorithm $L$ times a generator of $\gf{q^n}^\times$ can be distinguished from that of its subgroups with probability $> 1-\frac{1}{D^L}$.

At last, let us restrict our focus to finite fields with characteristic 2 ($q$ is a power of 2), in which case $\gf{q^n}$ can be represented with exactly $n\log q$ qubits. With $\ket{s}=\frac{1}{\sqrt{q^n}} H^{\otimes n\log q}\ket{\mathbf{0}}$ being the uniform superposition of all elements of $\gf{q^n}$ there is  $\braket{s|\psi_{n-1}}\approx\sqrt{r/(q^n-1)}$, especially,  $\braket{s|\psi_{n-1}(\alpha)}=\sqrt{q^n/(q^n+2)}$ when $\alpha$ is primitive.  The last two stages of the previous algorithm can be replaced by simply applying $H^{\otimes n\log q}$ to $\ket{\psi_{n-1}}$, then measurements on the $n\log q$ qubits in computational basis will result in a string of zeros with probability $q^n/(q^n+2)$ for primitive $\alpha$, whereas that probability is only approximately $r/(q^n-1)<1/D$ for any non-primitive $\alpha$.

\subsection{Complexity}\label{complexity}

The complexity of our first algorithm mainly comes from the multipliers $\{U_{x^{q^k}}\}_{k=0}^{n-1}$ for multiplication between elements of the finite field $\gf{q^n}$. Quantum multipliers over $\gf{2^n}$ have been studied in \mycites{beauregard03,maslov09,rotteler13,amento13,amento2013,kepley15,imana21}, typically with a gate count of $O(n^2)$ and a depth of $O(\log^2n)$. It is worth mentioning that a quadratic gate count may not be optimal and can be further reduced. In particular, the authors of  \mycite{kepley15} showed that a multiplication circuit with a gate count of  $O(n^{\log3})$ is available on condition that there exists an irreducible trinomial over $\gf{2}$ of degree $n$. As for quantum multipliers over general finite fields $\gf{q^n}$, considering the multipliers over characteristic-2 finite fields it is reasonable to expect a gate count of $O(n^2\log^2 q)$ and a depth of $O(\log^2 \log q^n)$.

In comparison to the direct application of Shor's algorithm to factor the integer $q^n-1$ followed by classical algorithms to assess the primitivity of degree-$n$ irreducible polynomials over $\gf{q}$, our algorithm makes full use of the fact that the order of a group is divisible by the order of its elements. This realizes an approximately 1/3 reduction in qubits and a 1/2 reduction in the number of gate operations in individual runs of the algorithm. To be specific, our approach requires only $n$ qubits in the first register, whereas Shor's algorithm necessitates 2$n$ qubits to control the $n$ qubits in register 2. Moreover, the methods for reducing qubits in order-finding algorithms developed in \mycites{zalka06,griffiths96,parker00,smolin2013oversimplifying} can be applied to our algorithm directly, but these simplifications do not result in a further reduction in the gate count.

Regarding the second algorithm, its complexity depends on the reversible quantum general multiplier (RQGM) and the quantum Fibonacci automorphism over $\gf{q^n}$. Here requiring the multipliers to be reversible would save considerable computation space, as the state $\ket{\psi_k}$ \eqref{psi} will be automatically initialized to $\ket{\mathbf{0}}$ immediately after $\ket{\psi_{k+1}}$ is obtained. Recently, the authors of  \mycite{imana21} proposed an optimized reversible quantum multiplier over $\gf{2^8}$ which consists of 64 Toffoli gates and approximately 15 CNOT gates and is implementable on 24 qubits. On the other hand, according to \myref{QFA}, the Fibonacci automorphism $U_A$ can be realized by multiplying $s_{k}\in\gf{q}$  and the polynomial remainder of $x^{qk}$ modulo $p(x)$ independently for each  $0\leq k\leq n-1$,  and then add up the $n$ resulting  polynomials. Thus $U_A$ can be factored  into $n$ parallelized multiplications between an element of $\gf{q}$ and a polynomial over $\gf{q}$, followed by $n$ parallelized additions between polynomials over $\gf{q}$. In this way, the gate count of $U_A$ is $O(n^2\log ^2q)$.

It is worth noting that the quantum circuits  \cite{beauregard03,maslov09,rotteler13,amento13,amento2013,kepley15,imana21} for implementing modular multiplication between degree-$n$ polynomials over $\gf{q}$, i.e., quantum multipliers between elements of $\gf{q^n}$, may be not be optimal. Optimizing these circuits will significantly  reduce the overall complexity of our algorithms, especially considering that most of the complexity comes from the modular multiplications. Low-energy implementations of modular multiplications as well as other arithmetic operations would of course make our algorithms more efficient in practical applications. Previous efforts to improve the efficiency of modular multiplications have been explored in \cite{Elkhatib2021,cintas-canto}. It will be intriguing to explore how similar approaches can be leveraged in the context of quantum modular multiplications.

\section{Discussions}\label{discussion}

Our algorithms have their respective pros and cons. Firstly, unlike the second algorithm, the first one introduces an extra computational cost during the initial stage because it requires the preparation of a uniform superposition of $q^n-1$ orthogonal states. Secondly, the quantum multipliers involved in the first algorithm can be implemented on exactly $2n$ qubits, whereas the second one requires additional ancillary qubits. Thirdly, unlike the modular exponential in order-finding algorithms,  the multipliers over finite fields in the second algorithm are not controlled by (entangled with) additional qubits. This lack of entanglement makes them more robust against decoherence. Fourthly, the presence of unavoidable phase noise in any physical system may significantly limit the accuracy of the Quantum Fourier Transform (QFT). The controlled phase rotations involved in the QFT, represented by matrices $P_i$ as given by
\begin{equation*}
P_i=\begin{pmatrix}
	1&0\\
	0&e^{i2\pi/2^k}
\end{pmatrix}\ (1\leq k\leq n),
\end{equation*}
could be completely obscured by noise, particularly for sufficiently large $k$.  In contrast, the second algorithm, especially when $q$ is a power of 2, exhibits stronger robustness against phase rotation noise. Fifthly, a notable drawback of the second algorithm is its inability to definitively determine whether an element of $\gf{q^n}$ is primitive. Nonetheless, the confidence level approaches 1 exponentially fast as we repeatedly implement the algorithm.

In cryptographic applications of primitive polynomials over finite fields,  such as code-based digital signature schemes \cite{kuznetsov19,gorbenko16,stern1993,veron1997,song2020} and Y00 protocol \cite{yuen03,yuen04,Barbosa03,Yuen09,chuang05}, it is crucial that the specific primitive polynomial, which are shared between authorized participants through secure secret keys, remains confidential to potential eavesdroppers. However, an eavesdropper may get side-channel information about the secret keys utilizing proper side-channel attacks, passively or actively. Thus these attacks are essential for the security analysis of protocols involving primitive polynomials. For a comprehensive analysis on implementation attacks haunting PQC,  we recommend \mycites{canto2023,ali2016,dubrova2023,cintas-canto2023}.

\section{Conclusion }\label{conclusion}

We propose two efficient quantum algorithms for generating random primitive polynomials over finite fields. The first one is essentially a simplified order-finding algorithm for assessing the primitivity of randomly selected irreducible polynomials. In contrast, the second one is a new algorithm which can be employed to obtain primitive elements of  $\gf{q^n}$. Based on each primitive element of $\gf{q^n}$, a unique degree $n$ primitive polynomials over $\gf{q}$ can be efficiently constructed on a classical computer.

Though being probabilistic, the second algorithm requires no quantum gates that are controlled by ancillary qubits, rendering it intrinsically more robust against decoherence. Moreover, it circumvents the intricate phase rotation fine-tuning required in order-finding algorithms. To leverage the complementary strengths of both algorithms, one can employ the second algorithm to generate random polynomials that are primitive with high probability, and then test their primitivity unambiguously with the first algorithm. This above scheme not only saves valuable quantum resources but also ensures a reliable assessment of primitivity, making it more suitable for practical applications.

Our research enables the rapid and real-time generation of random high-order primitive polynomials over finite fields, which lay the foundation for the sophisticated applications of primitive polynomials in diverse communication and computation tasks, such as code-based post-quantum cryptography, digital signature schemes, and many other quantum communication protocols.  In the future, one of our focus will be on optimizing the quantum multipliers for elements of finite fields.

\appendix

\section{\label{Togate}Multiple Control Toffoli Gate}

In order to prepare a uniform superposition of $N=2^n-1$ orthogonal states,  the following $n$-qubit controlled NOT gate is required:
$
\frac{1}{\sqrt{N+1}}\sum_{j=0}^{N-1}\ket{j}\ket{0}_{\rm anc}\stackrel{}{\longrightarrow}\\\frac{1}{\sqrt{N+1}}\big(\sum_{j=0}^{N-2}\ket{j}\ket{0}_{\rm anc}+\ket{1}^{\otimes n}\ket{1}_{\rm anc}\big).
$
The measurement result on the ancillary qubit will be 0 and 1 with probability $\frac{2^n-1}{2^n}$ and $\frac{1}{2^n}$ respectively. There has been plenty of investigations on multiple control Toffoli gates (multi-qubit controlled NOT gate) \cite{seilingerT-depth,multiplecontrolcavity,he2017decomposition,maslov2016toffoli}. Among them the Toffoli gates with relative phase are generally simpler \cite{maslov2016toffoli} and, since our algorithm is unaffected by any relative phase transform on the ancillary qubit, the best choice for us may be the one proposed in  \mycite{maslov2016toffoli}, which transforms the state $\ket{1}^{\otimes n}\ket{0}$ to $e^{i\pi/2}\ket{1}^{\otimes n}\ket{1}$ while keeping any other states invariant via $8n-17$ T gates, $6n-12$ CNOT gates, $4n-10$ Hadamard gates and, $\lceil \frac{n-3}{2}\rceil$ ancillary qubits initialized and returned to $\ket{0}$.

\vskip 0.5cm

\section{\label{QFT}Approximated Quantum Fourier Transform}

Perfect Fourier transform over $\mathbb{Z}_{N}$ generally can not be easily implemented when $N$ is not a power of 2, we  turn to algorithms for approximating Fourier transform with high accuracy and efficiency, which have been largely investigated \cite{ApproFourier,kitaev1995quantum,hales1999quantum,coppersmith1994approximate}.  Here we'd like to mention Hales and Hallgren's work on approximated quantum Fourier transform over arbitrary $\mathbb{Z}_N$  $(N\in N^+)$ \cite{ApproFourier}, where they propose a method to $\epsilon$-approximate the quantum Fourier transform over $\mathbb{Z}_N$ for any $N$ and any $\epsilon$ in time $O\big(\log N\log\frac{\log N}{\epsilon}+\log^2\frac{1}{\epsilon}\big)$ using $\log N+O(\log\frac{1}{\epsilon})$ qubits. At the same time, due to the unitary evolution of quantum computation the above polynomial approximation is sufficient for any efficient algorithm in practice \cite{SWOfQComputing}.

Specifically, when $N=2^n-1$ $(\log N\approx n)$, according to \cite{ApproFourier}, we are able to approximate a perfect Fourier transform over $\mathbb{Z}_{2^n-1}$ with accuracy $\epsilon$ on $n+O(\log \frac{1}{\epsilon})$ qubits
and in time  $O\big(n\log \frac{n}{\epsilon}+\log^2\frac{1}{\epsilon})$.\\

\noindent {\bf Acknowledgments}\\

\noindent This work is supported by the National Natural Science Foundation of China (Grants No. 12175104 and No. 12274223), the Innovation Program for Quantum Science and Technology (2021ZD0301701), the Natural Science Foundation of Jiangsu Province (No. BK20211145), the Fundamental Research Funds for the Central Universities (No. 020414380182), the Key Research and Development Program of Nanjing Jiangbei New Area (No. ZDYD20210101),  the Program for Innovative Talents and Entrepreneurs in Jiangsu (No. JSSCRC2021484), and the Program of Song Shan Laboratory (Included in the management of Major  Science and Technology Program of Henan Province) (No. 221100210800-02).\\

\noindent{\bf Conflict of Interest}\\

\noindent The authors declare no conflict of interest.\\

\noindent{\bf Keywords}\\

\noindent Quantum algorithm, primitive polynomials, finite fields, quantum cryptography


\end{document}